\documentclass[12pt]{iopart}
\usepackage{iopams}  
\usepackage{epsfig}

\newcommand{\bea}{\begin{eqnarray}}
\newcommand{\eea}{\end{eqnarray}}
\newcommand{\vect}[1]{\mathbf{#1}}

\newcommand{\req}{\rho_{\rm eq}}

\newcommand{\kt}{k_{\rm B}T}
\newcommand{\imag}{{\rm i}}
\begin{document}

\title{Mode expansion for the density profile of crystal--fluid interfaces: Hard spheres as a test case }

\author{M. Oettel}

\address{Johannes Gutenberg--Universit\"at Mainz, Institut f{\"ur} Physik,
  WA 331, D--55099 Mainz, Germany}
\ead{oettelm@uni-mainz.de}
\begin{abstract}
 We present a technique for analyzing the full three--dimensional density profiles   
 of a planar crystal--fluid interface in terms of density modes. These density modes
 can also be related to crystallinity order parameter profiles which are used in 
 coarse--grained, phase field type models of the statics and dynamics of crystal--fluid 
 interfaces and are an alternative to crystallinity order parameters extracted from
 simulations using local crystallinity criteria. We illustrate our results for the
 hard sphere system using finely--resolved, three--dimensional density profiles
 from density functional theory of fundamental measure type.
\end{abstract}

\maketitle

\section{Introduction}

The study of interfaces between the thermodynamically stable bulk phases continues to be of great practical and theoretical
interest. One main motivation is simply the fact that every finite sample of material in some environment contains interfaces. If the
bulk sample is sufficiently small, the physical properties of these interfaces may even govern the behavior of
the whole sample. Additionally, surface--specific processes are widely investigated in materials science.

The liquid--vapor interface appears to be the simplest of these interfaces. For a simple liquid, 
there is just one scalar order parameter, the one--particle density, which continuously changes when going from the vapor to the liquid.
The basic statistical mechanical theory of the liquid--vapor interface has already been developed a while ago
\cite{Eva79} but, of course, there are still open problems such as the question of the 
correct formulation of effective interface Hamiltonians and the behavior of the 
wavevector--dependent surface tension \cite{Mec99,Tar07}. 

The solid--liquid and solid--vapor interfaces appear to be  more difficult to describe. Upon solidifying, the continuous
translational symmetry of the liquid (vapor) is broken to a discrete symmetry describing the crystal.
Connected to this, besides density a number of additional order parameters are needed which completely 
characterize the bulk crystal and which should vanish when going over to the bulk liquid (vapor) phase. 
Phenomenological attempts to capture the liquid--solid transition from a point of view of statistical mechanics 
have restricted the choice of order parameters to density and one crystallinity parameter (models of phase
field type \cite{Hoy03}) and employed ad--hoc free energies of Landau--Ginzburg type to incorporate the transition. 
A more fundamental approach is  classical density functional theory (DFT) \cite{Eva79} in which all equilibrium properties
are governed by a unique free energy functional of the one--particle density. In this approach, the crystalline
order parameters can be related to the Fourier modes of the inhomogeneous density profile. Depending on
the level of approximation used, only a few modes are needed (as in phase field crystal models \cite{Emm11}) or     
many (as in Ramakrishnan--Yussouff models, weighted density functionals, \dots \cite{book:das} ). In practical calculations, however,
the density mode expansion is almost always severely truncated. 

From the DFT point of view, the full three-dimensional density profile of a solid--liquid interface 
contains all necessary information about the crystalline order parameter profiles. As remarked, with very few 
exceptions existing DFT calculations used restricted parametrizations and none of them attempted a systematic analysis
of the three-dimensional density profile in terms of modes. In this paper, we present such a mode analysis of
the full density profile which is general and the resulting modes are also quantities which are (in principle) observable 
in scattering experiments or real--space experiments (confocal microscopy). We exemplify the technique using very 
accurate DFT data (using fundamental measure functionals) for the hard sphere crystal--liquid interface.
It appears also to be interesting to analyze simulation data involving solid--liquid interfaces using density modes.
Quite often, effective order parameter profiles are measured in simulations through bond--order variables which are,
however, difficult to capture in theory as they are connected to $n$--point correlation functions where
$n$ is the number of next neighbours of a particle or a multiple thereof.  

The paper is structured as follows: In Sec.~\ref{sec:mode_analysis} we introduce the mode analysis in general terms.
Sec.~\ref{sec:results} contains results of the mode analysis of our DFT data. Here we briefly introduce the fundamental
measure functionals, review the thermodynamic observables for the bulk crystal and the crystal--liquid interfaces and
then discuss the density as well as the free energy modes. Sec.~\ref{sec:summary} contains a discussion on the 
meaning and measurability of the modes and some conclusions.

\section{Mode analysis}

\label{sec:mode_analysis}

Consider the density distribution $\rho_{\rm cr}(\vect r=(x,y,z))$ in a perfect crystal. It can be Fourier expanded
into a discrete sum,
\bea
  \rho_{\rm cr}(\vect r) = \sum_j P_j\;\exp(\imag \vect K_j \cdot \vect r)\; ,
\eea
where $\vect K_j$ denotes the set of all reciprocal lattice vectors (RLV) and $P_j$ are the associated Fourier amplitudes.
Owing to the symmetry of the crystal under consideration, not all $P_j$ are
independent of each other. It is convenient to group the $\vect K_j$ in shells with index $m$
where all $\vect K_j$ belonging to one shell can be transformed into each other under the discrete symmetry group
of the crystal under consideration (and thus all $P_j$ associated with these $\vect K_j$ are equal).
As an example, for the face-centered cubic ({\em fcc}) crystal the reciprocal lattice is of body-centered cubic ($bcc$)
symmetry. Let $a$ be the side--length of the cubic unit cell of {\em fcc} and correspondingly $b=2\pi/a$ the
side length of the cubic unit cell of $bcc$ in reciprocal space. The reciprocal basis is given 
(in Cartesian coordinates where the axes span the cubic unit cell in reciprocal space) by
$\vect B_1 = b(1,1,-1)$, $\vect B_2 = b(1,-1,1)$ and $\vect B_3 = b(-1,1,1)$. An arbitrary RLV is a linear combination
of the $\vect B_i$. The shells are characterized by a triple $(m,n,k)$ of natural numbers and the $K_j$
belonging to this shell have Cartesian components $b(\pm m, \pm n, \pm k)$ and permutations thereof. Thus, if
$m,n,k$ are mutually distinct, there is a maximum of 48 RLV in one shell. The shells with lowest modulus are given 
by $(1,1,1)$, $(2,0,0)$ and $(2,2,0)$. A listing of the RLV triples up to shell 15 is given in Ref.~\cite{Hay83} (Table I).  

\begin{figure}
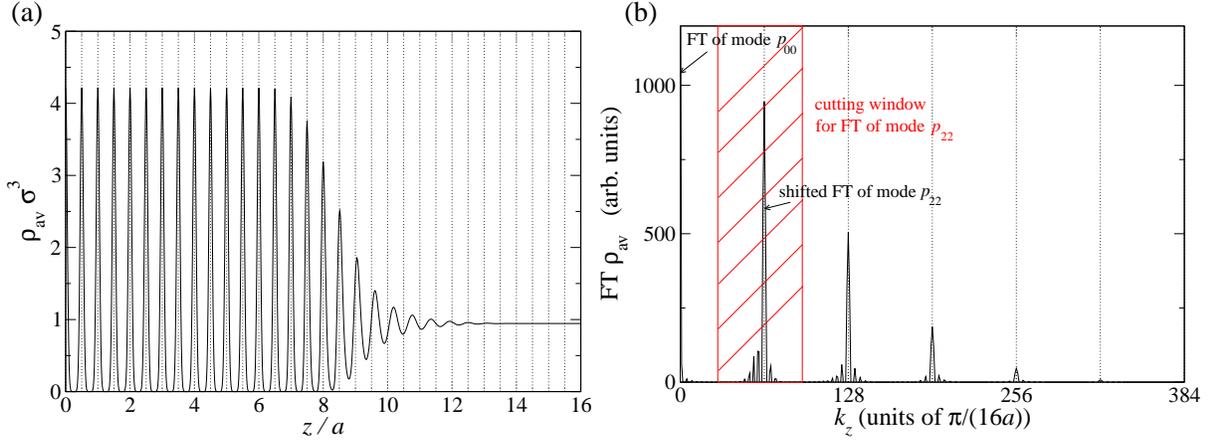

 \vspace*{1cm}
 \epsfig{file=zaverage_100.eps, width=7.8cm}
 \epsfig{file=FTrho_zaverage_100.eps, width=8cm}
 \caption{ (color online) Hard sphere crystal--fluid interface in [100] orientation. (a) Laterally averaged density profile. 
(b) Fourier transform of laterally averaged density profile. Clearly visible are the ``lumps'' around 
multiples of the wave numbers $4\pi/a$ (corresponding to the periodicity $a/2$ in the density profile). The back transforms
of the individual lumps define the mode profiles. 
 }
 \label{fig:zav1}
\end{figure}

We prepare a planar crystal--fluid interface with interface normal in $z$--direction and density
profile $\rho(x,y,z)$. It is tempting to parameterize the density profile using the RLV as before,
\bea
 \rho(x,y,z) = \sum_j \exp(\imag \vect K_j \cdot \vect r)\; p_j(z)\;,
\eea  
but now with a $z$--dependent Fourier amplitude $p_j(z)$ for the reciprocal modes.
We call them (density) modes or order parameter profiles.  
 One expects that upon
crossing the interface coming from the crystal side, all $p_j(z)$ go to zero (for nonzero $\vect K_j$). Only
for $\vect K_j \equiv 0$ the value for the associated mode crosses from the average density of the crystal to the average density of the
fluid. As discussed before, in the bulk crystal all RLV within one shell with index $m$ are degenerate.  
At the interface the
degeneracy of the RLV in one shell is lifted, and we introduce an index $n$ which
distinguishes the possible values of the $z$--component of the RLV. Thus the decomposition becomes
\bea
 \label{eq:int_deco} 
  \rho(x,y,z) = \sum_{mn} \sum_j p_{mn}(z)  \exp\left(\imag (\vect K_j)_{mn} \cdot \vect r\right)\;.
\eea
The sum over $j$ is now only for those RLV within shell $m$ which have common value of $z$--component, as expressed
by the index $n$. In the literature, such a decomposition has been used to parameterize the full 3d density profile
$\rho(x,y,z)$ using  only the leading mode \cite{Loe90,She96, Wu06} in order to facilitate a simplified
order parameter description of the crystal--fluid interface. Consistency with the bulk crystal solution
$\rho_{\rm cr}$ requires $p_{mn}(z \to -\infty) = P_m$, thus in a leading--mode description either the crystal
description is also truncated \cite{Wu06} or higher--than--leading modes are parametrized proportional
to the leading mode \cite{She96}.  

For a given  3d density profile $\rho(x,y,z)$ the extraction of the modes $p_{mn}^j(z)$ does not appear to be 
straightforward since they can not be projected out using a Fourier integral. 
Consider the Fourier transform of the 3d density profile:
\bea
 \label{eq:FTrho}
 \tilde\rho(k_x,k_y,k_z) = L_xL_y \sum_{mn}\sum_j  \delta^2 \left(\vect k_{\perp} +  (\vect K_{\perp,j})_{mn}\right)
 \tilde p_{mn} \left(k_z + (K_{z})_{mn}\right)\; .
\eea
Here, the lateral dimensions of the system with interface are denoted by
$L_x$ and $L_y$.
The wavevector in the interface plane, $k_\perp=(k_x,k_y)$, is restricted to the discrete values
allowed by the RLV, $(\vect K_{\perp,j})_{mn}$. The Fourier transform of the modes, $\tilde p_{mn}$,
are shifted by the $z$--component of the associated RLV, $(K_{z})_{mn}$, which is by our convention
fully specified by the indices $m,n$. Hence we see that the $\tilde p_{mn}(k_z)$ can be viewed as
being centered around $-(K_{z})_{mn}$ and are well separable from each other 
as long as their range in Fourier space does not extend further than the separation between two neighbouring
$(K_{z})_{mn}$. This in turn  is given by $2\pi/d_{[o]}$ with $d_{[o]}$ being the $z$--separation of lattice planes 
in the orientation [$o$] . This condition on the mode separability is a very physical one:
it means that the density modes should vary across the interface much more smoothly than the actual density profile
$\rho(x,y,z)$.   

We illustrate the mode separability using explicit DFT results for the crystal--fluid interface in the hard sphere system,
for more details see Sec.~\ref{sec:results} below. For the [100] orientation, Fig.~\ref{fig:zav1} (a) shows the 
laterally averaged density profile 
\bea
  \rho_{\rm av}(z) = \frac{1}{L_x L_y} \int_0^{L_x} dx \int_0^{L_y} dy\; \rho(x,y,z) \;,
\eea   
and Fig.~\ref{fig:zav1} (b) its Fourier transform, corresponding to $\tilde\rho(0,0,k_z)$. 
This Fourier transform consists of localized ``lumps'' around the values $i \cdot 4\pi/a$ $(i =0,1,2,\dots$) 
which mark lattice planes separated by the distance $d=a/2$ (which is what we expect for the laterally averaged 
density of an {\em fcc} crystal). The inverse Fourier transform of the isolated ``lumps'' (centered around 0) gives the 
associated modes in real space. As indicated graphically in Fig.~\ref{fig:zav1} (b), the Fourier transform
of a particular mode can be isolated by cutting out the Fourier transform $\tilde\rho(0,0,k_z)$ symmetrically 
around $i \cdot 4\pi/a$ with a width $4\pi/a$. We formalize this a bit more generally as
\bea
 \label{eq:pmn_extract}
  \tilde p_{mn}(k_z) = \tilde \rho( K_x, K_y, k_z+(K_{z})_{mn}) \; g(k_z)
\eea  
where $(\vect K_{\perp,j})_{mn}=(K_x,K_y)$ denotes the in--plane components of the RLV of interest for
mode $(mn)$ (again, there is an additional degeneracy, expressed by $j$). The function $g(k_z)$ is a suitable cutoff function,
varying monotonically from 1 at around $k_z=0$ to zero near $k_z={2\pi}/{d_{[o]}}$
and with the additional properties
\bea
 \label{eq:gk_def}
   g(k_z) &=& 0 \qquad  \left(|k_z|>\frac{2\pi}{d_{[o]}}\right) \;,  \nonumber \\
   g(k_z) -g(-k_z)  &=& 0 \;, \label{eq:gkconds} \\ 
   g(k_z) + g\left(\frac{2\pi}{d_{[o]}}-k_z\right) &=& 1 \qquad \left(k_z \in \left[0, \frac{2\pi}{d_{[o]}}\right]\right) \;. \nonumber
\eea
Note that these conditions are required to have (\ref{eq:pmn_extract}) consistent with (\ref{eq:FTrho}).
A practical choice (which violates the first condition a tiny bit) is
\bea
  g(k_z>0)  = \frac{1}{2}\left(1 - \mbox{erf}\left(\alpha\,\left[\frac{k_z d_{[o]}}{\pi}-1\right]\right) \right) 
\eea 
where $\alpha$ tunes the width of the error function kink. Note that 
for $\alpha\to\infty$ the cutoff function becomes $g(k_z>0) = \theta(\pi/d_{[o]}-k_z)$, literally corresponding to
the graphical procedure indicated in Fig.~\ref{fig:zav1} (b). The disadvantage hereby is
that through a hard cutoff in Fourier space unwanted oscillations arise in real space which simply might be
due to insufficient resolution of the density profile in real space and are not due to physical oscillations
of the modes.
We found $\alpha > 3$ a convenient choice. 

We had introduced the modes $p_{mn}(z)$ as belonging to the RLV shell $m$ but differing in
the $z$--component of the associated RLV, as indicated by $n$. If that $z$--component is zero, 
the mode will be purely real. If for a particular $n$ that $z$--component is nonzero, the mode will be in general complex
since from Eq.~(\ref{eq:pmn_extract}) it can not be expected that the Fourier transform of the mode
is symmetric in $k_z$. The reality of $\rho(x,y,z)$ requires that in the same shell $m$ there is
a mode $p_{m\bar{n}}(z)$ for which the $z$--component of the associated RLV is minus the previous one and which is the 
complex conjugate of $p_{mn}(z)$.
Hence can define real--valued modes through separating real and imaginary part,
\bea
 \nonumber
  p_{mn}^+ (z) = \frac{p_{mn}(z)+p_{m\bar{n}}(z)}{2}\;, & \qquad & 
  p_{mn}^- (z) = \frac{p_{mn}(z)-p_{m\bar{n}}(z)}{2\imag} \;,
\eea
and the $p_{mn}^- (z)$ have the obvious interpretation of {\em phase shifts} of the associated density mode
oscillations across the interface. 

\section{Results for the hard sphere crystal--fluid interface}
\label{sec:results}

\subsection{Fundamental measure DFT}

We apply the mode expansion technique introduced in the previous section to DFT results for the
full crystal--fluid interface density profile. DFT is built on the existence of a unique free energy 
functional of the one--particle
density field $\rho(\vect r)$,
\bea
  {\cal F}[\rho] & = &{\cal F}^{\rm id}[\rho] + {\cal F}^{\rm ex}[\rho] \;, \\
  \label{eq:fid}
   \beta  {\cal F}^{\rm id}[\rho] &=& \int d^3 r \rho(\vect r)\left( \ln(\rho(\vect r)\Lambda^3) -1 \right)         
\eea
which can be split into the exactly known ideal gas part ${\cal F}^{\rm id}$ ($\Lambda$ is the de--Broglie wavelength,
$\beta=1/\kt$ is the inverse temperature)
and a generally unknown
excess part ${\cal F}^{\rm ex}$. The equilibrium density $\req(\vect r)$ in the presence of an external
(one--particle) potential $V^{\rm ext}(\vect r)$ is then given by
\bea
 \label{eq:minimizingF}
  \left.\frac{\delta {\cal F}[\rho]}{\delta\rho(\vect r)}\right|_{\rho=\req} = \mu - V^{\rm ext}(\vect r)\;,
\eea
where $\mu$ is the imposed chemical potential (e.g. by requiring a certain bulk density far away from the
region where the external potential acts). For the description of the equilibrium density profile between
crystal and fluid, $V^{\rm ext}=0$. Thus the bulk crystal far away from the interface on one side appears
as a self--sustained, periodically inhomogeneous fluid at the coexistence chemical potential $\mu_{\rm coex}$.  

For the hard sphere system, fundamental measure theory (FMT) allows the construction of very precise functionals
\cite{Ros89,Tar00,Rot10}.
Essentially, FMT postulates an excess free energy with a  local free energy
density in a set of weighted densities $n_\alpha$:
\bea
  {\cal F}^{\rm ex}[\rho] &=& \beta^{-1} \int d^3r \Phi(n_\alpha(\vect r)) \;.
\eea
The weighted densities are constructed as convolutions of the density with weight functions,
$n_\alpha(\vect r) = \rho * w^\alpha(\vect r)$.
The weight functions reflect
the geometric properties of the hard spheres.
For one species, the weight functions include four scalar functions
$w^0 \dots w^3$, two vector functions $\vect w^0, \vect w^1$ and a tensor function
$w^t$ defined as
\bea
 w^3 = \theta(R-|\vect r|)\;, \qquad
 w^2 = \delta(R-|\vect r|)\;, \qquad w^1 = \frac{w^2} {4\pi R}\;,
  \qquad w^0 = \frac{w^2}{4\pi R^2}\;,
  \nonumber\\
 \vect w^2 =\frac{\vect r}{|\vect r|}\delta(R-|\vect r|)\;,
 \qquad \vect w^1 = \frac{\vect w^2}{4\pi R} \;, \nonumber \\
 w^t_{ij} = \frac{r_i r_j}{\vect r^2} \delta(R-|\vect r|) \;.
\eea
Here, $R=\sigma/2$ is the hard sphere radius. Using these weight functions,
corresponding scalar weighted densities $n_0\dots n_3$, vector weighted densities
$\vect n_1, \vect n_2$ and one tensor weighted density $n_t$ are defined.
In constructing the free energy density $\Phi$, arguments
concerning the correlations in the bulk fluid, certain geometric consistencies and 
arguments pertaining to strongly inhomogeneous
systems are used  \cite{Rot10}:
\bea
 \label{eq:phi_hs}
 \fl  \Phi( \{\vect n[\rho (\vect r)]\} ) &=&   -n_0\,\ln(1-n_3) +
      \varphi_1(n_3)\;\frac{n_1 n_2-\vect n_1 \cdot \vect n_2}{1-n_3} +
   \nonumber   \\
      & & \varphi_2(n_3)\; \frac{3 \left( -n_2\, \vect n_2\cdot \vect n_2 + n_{2,i} n_{t,ij} n_{2,j}
        + n_2\,  n_{t,ij}  n_{t,ji} - n_{t,ij}  n_{t,jk} n_{t,ki}
   \right) }{16\pi(1-n_3)^2}\;.
\eea
Here, $\varphi_1(n_3)$ and $\varphi_2(n_3)$ are functions of the local packing density $n_3(\vect r)$.
The most accurate functional to date is the tensor version of the recently introduced 
White Bear II (WBII) functional \cite{Han06} and corresponds to the choice
\bea
 \label{eq:fwbII}
 \varphi_1 & = & 1 + \frac{2n_3-n_3^2 + 2(1-n_3) \ln(1-n_3 )}{3n_3} \\
 \varphi_2 & = & 1 - \frac{2n_3-3n_3^2 + 2n_3^3 + 2(1-n_3 )^2 \ln(1-n_3 ) }
                          {3 n_3^2} \;.  \nonumber
\eea
This functional has proved to be very reliable for strongly inhomogeneous fluids \cite{Oet09,Bot09} and also
wall--fluid surface tension data at densities close to freezing excellently compare with
recent simulation data \cite{Deb11}. 

\begin{table}[h]
\begin{center}
\begin{tabular}{lllllll}
   \hline \hline
        &  $\rho_{\rm cr}\sigma^3$ & $\rho_{\rm fl}\sigma^3$ & $(\beta F/N)_{\rm cr}$ & $\beta\mu_{\rm coex}$ & $\beta p_{\rm coex}\sigma^3$ & $n_{\rm vac}$ \\ \hline \hline
FMT     & 1.039 & 0.945 & 4.96 & 16.38 & 11.87 & $2 \cdot 10^{-5}$ \\
SIM     & 1.041$^1$ & 0.940$^1$ & 4.96$^2$ & 16.09$^3$ & 11.58$^1$ & $3 \cdot 10^{-4}$$\quad^4$ \\ \hline \hline
        & $\beta\gamma_{\rm [100]}\sigma^2$ &  $\beta\gamma_{\rm [110]}\sigma^2$ & $\beta\gamma_{\rm [111]}\sigma^2$ &
          $\beta\gamma_{\rm [211]}\sigma^2$ & $\beta\gamma_{\rm [210]}\sigma^2$ & \\ \hline \hline
FMT$^5$   & 0.69 & 0.67 & 0.64 & 0.65 & 0.67 &   \\
SIM$^5$ &  0.64 & 0.62 & 0.60 & 0.61 & 0.62 & \\ \hline \hline
\end{tabular}
\end{center}
\begin{flushleft}
$^1$ From Ref.~\cite{Zyk10}. \\
$^2$ Free energy for $\rho_{\rm cr}\sigma^3=1.041$ \cite{Zyk10} using an improved fit in the form of the Speedy
equation of state from Ref.~\cite{Oet10}. \\
$^3$ Chemical potential for $\rho_{\rm fl}\sigma^3=0.940$ \cite{Zyk10} from the Carnahan--Starling equation of state.\\
$^4$ From Ref.~\cite{Kwa08}. \\
$^5$ From Ref.~\cite{Har12}.
\end{flushleft}

\caption{Thermodynamic properties of hard sphere crystals and crystal--fluid interface from FMT in comparison to 
simulation data.}
\label{tab:cr}
\end{table}

\subsection{Thermodynamic properties of the crystal and the crystal--fluid interface}

In equilibrium, hard spheres in their crystalline state form  an {\em fcc} lattice, although the free energy difference
to {\em hcp} is very small, about $10^{-3}$ $\kt$ per particle \cite{Koch05}. Bulk coexistence properties 
(densities, chemical potential, pressure, free energy) have been widely investigated by simulations and there seems to be consensus
on the respective values with an error margin below 1\%. The bulk crystal has been investigated in Ref.~\cite{Oet10} using
full minimization of the FMT functional discussed above and very good agreement has been found with simulations. The
coexistence density on the liquid side is overestimated by 0.5\%, this results in an overestimation of the
coexistence chemical potential and pressure of about 2\%. The relative vacancy concentration $n_{\rm vac}=2\cdot 10^{-5}$ 
at coexistence is too small by a factor of 10 compared to simulations, however, an assessment of possible systematic errors in
the simulations is not available. The FMT results for $n_{\rm vac}$ improve significantly on results from Ramakrishnan--Yussouff
and weighted density functional theories where $n_{\rm vac} \sim 0.1$. Also, in contrast to these earlier approaches,
the FMT crystal density profiles agree with simulations with respect to the width of the crystal peak and the
anisotropy in different lattice directions \cite{Oet10}. See Table~\ref{tab:cr} for a summary.  

The determination of crystal--fluid surface tensions in simulations has not yet resulted in consistent data 
across the literature. As an example, for the surface tension in [110] direction integration methods  
give $\beta\gamma_{\rm [100]}\sigma^2=0.58$ (cleaving, Ref.~\cite{Dav10}) and 0.64 (umbrella sampling, Ref.~\cite{Fer11}). Through the
capillary wave method, interfacial {\em stiffnesses} are accessible but the conversion to {\em tensions} is not
without difficulties owing  to the finite cutoff in the stiffness and tension expansion in cubic harmonics \cite{Har12}. 
Here, Laird et al. find $\beta\gamma_{\rm [100]}\sigma^2=0.58$ (expansion up to order two, Ref.~\cite{Dav06}) and 
newest results give 0.64 (expansion up to order three, Ref.~\cite{Har12}). The FMT results 
(also from Ref.~\cite{Har12}) are larger by about 7\%  and  the anisotropies compare well to the simulation results,
see Table~\ref{tab:cr}. It is perhaps not so surprising that the surface tension in FMT is higher than in simulations
since the FMT functional does not capture very long--ranged fluctuations present in capillary waves. Note that
previous theories based on ($i)$ severe approximations to the free energy functional and ($ii$) 
very restricted minimizations in density space have produced similar results \cite{Cur87,Mar93}, 
consequently the agreement in numbers has to be regarded as fortuitious. See also below for the discussion of the density modes
which puts simple density parameterizations in perspective.

\subsection{Density modes}

\begin{figure}
 \centerline{\epsfig{file=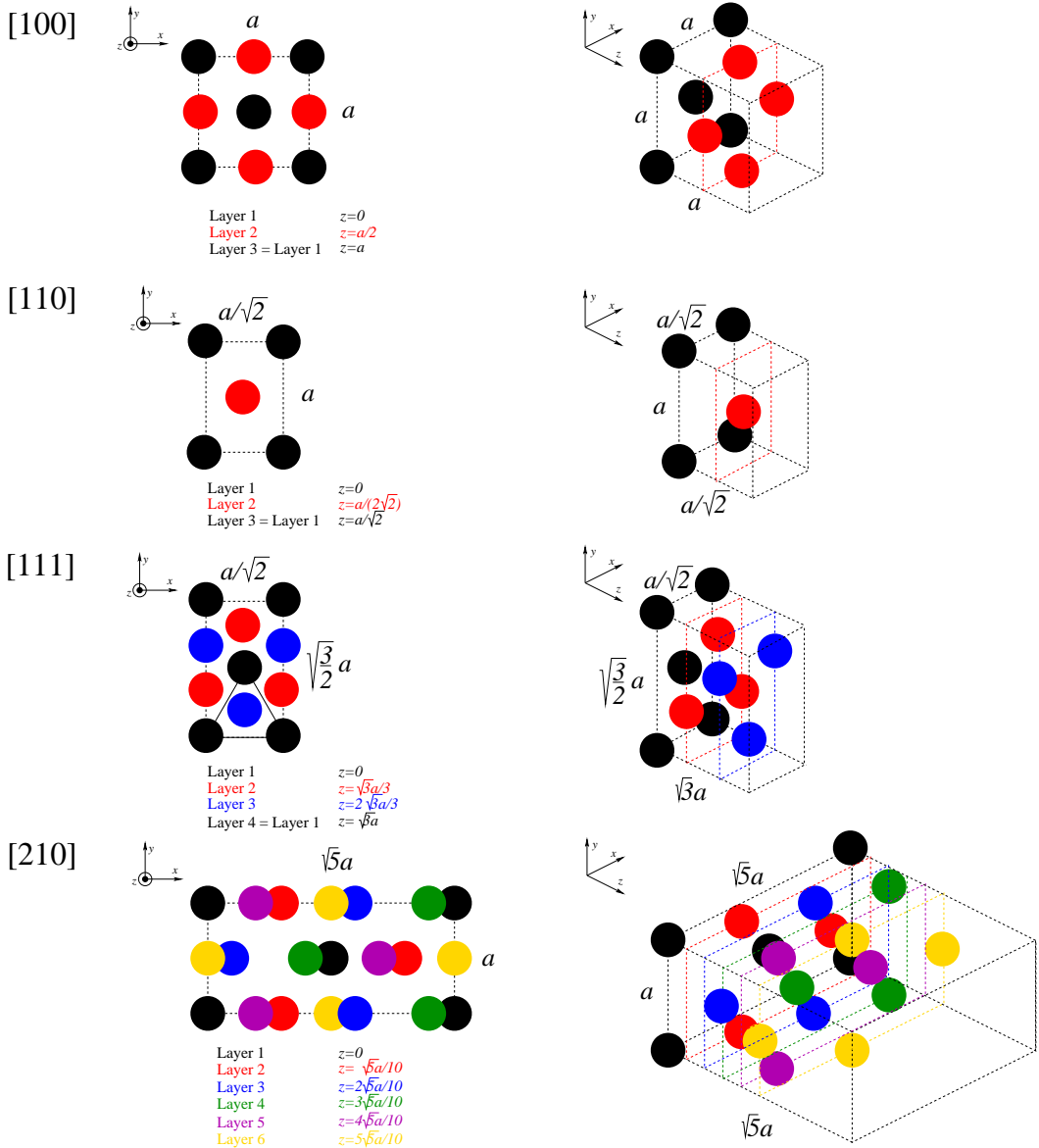, width=14cm}}
 \caption{(color online) Cuboid unit cells for different orientations of the interface normal ($z$--direction). Different layers with
  $z={\rm const.}$ are given in different colours. For [210], layers $n=7 \dots 11$ are omitted for clarity. They can be obtained 
  from layer $k=12-n$ using the reflection of coordinates $x \to \sqrt{5}a-x$ for all particles in the layer.
  The unit cell for the [211] orientation is obtained from [111] using the axis relabeling $x\to y$, $y\to z$ and $z\to x$.
  The $z$--separation between lattice planes, needed in Eqs.~(\ref{eq:pmn_extract}) and (\ref{eq:gk_def}), 
  is given by $d_{[100]}=a_z/2=a/2$,
  $d_{[100]}=a_z/2=a/(2\sqrt{2})$, $d_{[111]}=a_z/3=a/\sqrt{3}$, $d_{[211]}=a_z/6=a/(2\sqrt{6})$ and
  $d_{[210]}=a_z/10=a/(2\sqrt{5})$.
}
 \label{fig:faces}
\end{figure}
 
We analyze the density profiles of crystal--fluid interfaces in five different orientations of the interface
normal: [100], [110], [111], [211] and [210]. The density is given for a cuboid with edge lengths $L_x$, $L_y$
and $L_z$ which contains the fluid in the middle ($z\sim L_z/2$) and the crystal phase at the boundaries
($z \sim 0$ and $z \sim L_z$). $L_{x[y]}$ are given by the edge lengths in $x[y]$--direction, $a_{x,[y]}$, 
of the smallest cuboid unit cell of the crystal
which has the desired orientation in $z$--direction. Typically we chose $L_z=32 a_z$. The crystal cuboid unit cells
are depicted in Fig.~\ref{fig:faces}. Whereas they are still comparatively easy to imagine for the
[100], [110], [111] and [211] orientations, it is less trivial for the [210] orientation.   

As explained in Sec.~\ref{sec:mode_analysis}, the density modes $p_{mn}(z)$ are labeled by their RLV shell index $m$ and
by $n$ for the possibly different $z$--components of the RLV. The specific RLV assignment within one shell for the index $n$ then depends
on the chosen interface orientation. As an example, we show the assignment for the [100] interface in Table \ref{tab:pmn_100}.  
We call the mode $p_{11}(z)$ the leading crystallinity mode since the associated RLV is in the lowest shell and in the bulk crystal
the associated Fourier amplitude $P_1$ is largest. The meaning of $p_{11}(z)$ is linked to the strength of density oscillations
of close--packed planes across the interface into the bulk liquid. For the [100] interface, the normal to the close--packed planes
is {\em not} in $z$--direction. Another mode of significance is the mode $p_{22}(z)$ which is linked to the strength of density oscillations
of square arrays of particles which are in the $x$--$y$--plane. It is the leading Fourier component of the lateral density average
(often shown in simulation works) when the average density mode $p_{00}(z)$ has been subtracted.

\begin{table}
 \begin{center}
 \begin{tabular}{rrrr|rrr|l}
  \hline \hline
  $\quad m$ & $\quad n$ & $\qquad\qquad j$ & $\quad$ & $\quad K_x$ & $\quad K_y$ & $\quad K_z$ & {Significance} \\
    & & & & \multicolumn{3}{c|}{units of $2\pi/a$} &  \\ \hline
  0 & 0 &  1 && 0 & 0 & 0 & {average density} \\ \hline
  1 & 1 &  1\dots 4 && $\pm$ 1 & $\pm$ 1 &  1 & {leading crystallinity} \\ 
  1 & $\bar 1$ &  1 \dots 4 && $\pm$ 1 & $\pm$ 1 &  $-1$ & {mode} \\ \hline
  2 & 1 &  1\dots 2 &&  0 & $\pm$ 2 & 0 \\
    &   &  3\dots 4 &&  $\pm$ 2 & 0 & 0 \\
    & 2 &  1 &&  0 & 0 &  2 & {leading cryst. mode}  \\
   & $\bar 2$   & 1  && 0 &  0    & $-2$   &          {lateral density average} \\ \hline
  3 & 1 &  1\dots 4 && $\pm$ 2 & $\pm$ 2 & 0 \\
    & 2 &  1\dots 2 && 0 &  $\pm$ 2 & 2 \\
    &   &  3\dots 4 &&  $\pm$ 2 & 0 & 2 \\ 
    & $\bar 2$ &  1\dots 2 && 0 &  $\pm$ 2 & $-2$ \\
    &   &  3\dots 4 &&  $\pm$ 2 & 0 & $- 2$ \\ \hline \hline
 \end{tabular}
 \end{center}
 \caption{ The assignment of reciprocal lattice vectors to the main shell index $m=0 \dots 3$ and $z$--component index $n$ for the
 [100] interface. Note that modes with index $n$ and $\bar n$ are related by complex conjugation.   }
 \label{tab:pmn_100}
\end{table}

\begin{figure}
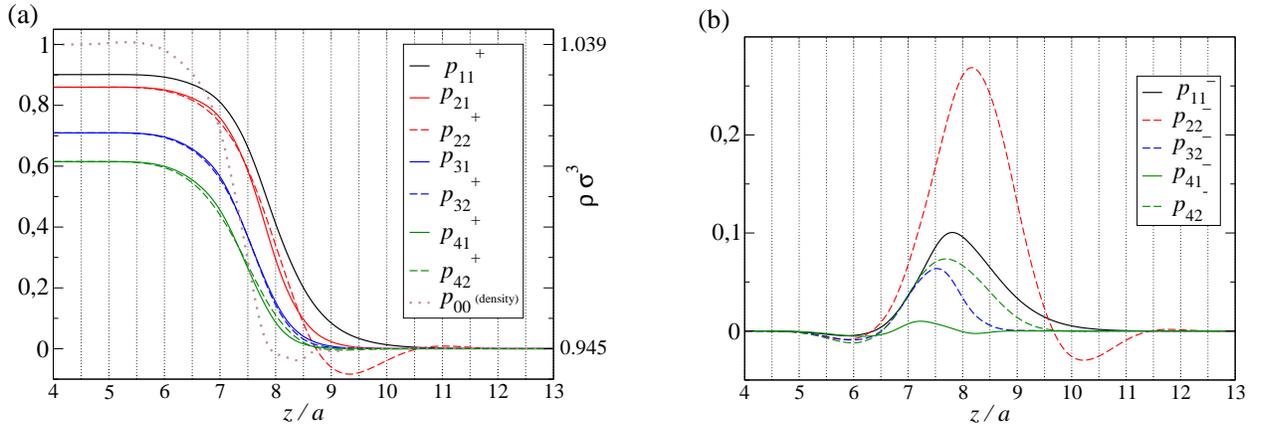

 \vspace*{1cm}
 \epsfig{file=rho_pmn_plus.c3.eps, width=8cm} \hspace{1.0cm}
 \epsfig{file=rho_pmn_minus.c3.eps, width=7.3cm}
 \caption{(color online) Density mode profiles $p_{mn}(z)$ in units of $\sigma^{-3}$  up to $m=4$ for the [100] interface. 
  (a) Real part of profiles, note that the density mode $p_{00}(z)$ has been rescaled and shifted, 
  see tick labels at right $y$--axis.
  (b) Imaginary part of profiles.  }
 \label{fig:modes_100}
\end{figure}

In Fig.~\ref{fig:modes_100} we show the modes up to $m=4$ for the [100] interface. 
In Fig.~\ref{fig:modes_100} (a), the real part of the mode profiles is shown, including the 
rescaled density mode $p_{00}(z)$ (fat dots). The density mode shows a depletion zone in front of the crystalline phase.
In previous work on the crystal growth at walls \cite{San11},  a depletion zone has been found in instantaneous profiles and
attributed to the {\em finite velocity} with which the interface moves. The absence of the depletion zone for {\em equilibrium} profiles
as seen in Ref.~\cite{San11} can be attributed to the averaging procedure: there the density was averaged between minima of the
laterally averaged density profile. Due to the phase shift of the density oscillations across the interface, such an averaging
procedure does not give our density mode $p_{00}(z)$; we verified that using the procedure of Ref.~\cite{San11}, also the
depletion zone vanishes. Further inspection of Fig.~\ref{fig:modes_100} (a) shows that the leading crystallinity mode
$p_{11}(z)$ shows a smooth, monotonic behavior which corresponds to the expectations on a phase--change order parameter profile and
which could be used to define an interface location. However, this obviously involves some arbitrariness: 
the kinks in higher modes $m\ge 2$ are clearly positioned deeper in
the crystal phase and the mode $p_{22}(z)$ (leading crystallinity mode of the lateral density average)
differs also in the qualitative shape by showing a 
local minimum (similar to the depletion zone in $p_{00}(z)$). 
Furthermode we note that the difference between the interface location as extracted from the leading crystallinity mode
$p_{11}(z)$ and the density mode $p_{00}(z)$ differ by about one unit cell length $a$. We discuss this further below.
-- In Fig.~\ref{fig:modes_100} (b) we show the imaginary part of the mode profiles, corresponding to the phase shift
of oscillations across the interface. Positive values indicate that the distance between maxima becomes larger
in the interface region, see also 
Fig.~\ref{fig:zav1} (a) where this is clearly visible for the laterally averaged density profile.

\begin{figure}[t]
 \vspace*{1cm}
 \centerline{\epsfig{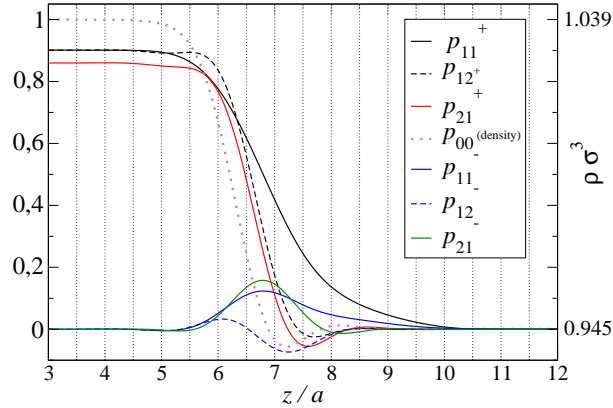} }
 \caption{(color online) Density mode profiles $p_{mn}(z)$ in units of $\sigma^{-3}$  up to $m=2$ for the [111] interface. 
  Note that the density mode $p_{00}(z)$ has been rescaled and shifted, 
  see ticks at right $y$--axis. }
 \label{fig:modes_111}
\end{figure}

As it is the case for the surface tensions, also the modes are not rotationally invariant. For  a new orientation,
one takes the rotation matrix which transforms the coordinate system of the [100] unit cell to the
cuboid unit cell with the desired orientation (see Fig.~\ref{fig:faces}) and applies it to the RLV.
The RLV within a shell $m$ need to be regrouped with respect to their new $z$--components
(see Table~\ref{tab:pmn_all_orient}). Whenever a $z$--degeneracy is lifted when going to another
orientation (such as for the leading mode $m=1$ which
is completely degenerate in [100] orientation), there will be 
new modes.
As an example, in Fig.~\ref{fig:modes_111} we show the modes up to $m=2$, both real and imaginary part,  
for the [111] interface. The shell $m=1$ is split into two modes where the associated RLV in reduced units
are given by $(0,0,3)$ (for $p_{11}(z)$) and $(0,2,1)$ (for $p_{12}(z)$). (The reduced units are defined
by normalization of RLV component $x[y,z]$ to the value $2\pi/a_{x[y,z]}$ where $a_{x[y,z]}$ are the side
lengths of the cuboid unit cell, see Fig.~\ref{fig:faces}.) As can be seen in Fig.~\ref{fig:modes_111}, 
mode $p_{11}(z)$ is similar in shape to the same mode for [100] and corresponds to the decay of the density oscillations
of the close--packed $x$--$y$--planes. However, the decay of density oscillations of 
the other close--packed planes which are {\em not}
perpendicular to the interface normal is different, as the behavior of $p_{12}(z)$ shows.  
 
\begin{table}
 \begin{center}
 \begin{tabular}{rr|rrrr|rrrr|rrrr|rrrr}
  \hline \hline
  $m$ & $n$ & $K_x$ & $K_y$ & $K_z$ && $K_x$ & $K_y$ & $K_z$ && $K_x$ & $K_y$ & $K_z$ && $K_x$ & $K_y$ & $K_z$ &    \\
      &     & \multicolumn{4}{c|}{[110]} & \multicolumn{4}{c|}{[111]} & \multicolumn{4}{c|}{[210]} & \multicolumn{4}{c}{[211]}   \\ \hline
   1 & 1 & $\pm$1 & $\pm$1 & 0      && $\pm$1 & 1      & $-$1   && $-$3 & $\pm$1 & $-$1 && $\pm$3 & 0      & 0 \\ 
     &   &        &        &        &&      1 & $\pm$1 &    1   &&    3 & $\pm$1 &    1 &&        &        &   \\
     &   &        &        &        &&      0 & $-$2   & $-$1   &&      &        &      &&        &        &   \\
     &   &        &        &        &&      0 & 2      &    1   &&      &        &      &&        &        &   \\
   1 & 2 &     0  & $\pm$1 & $\pm$1 &$^*$&      0 &    0   & $\pm$3 &$^*$&    1 & $\pm$1 & $-$3 &$^*$& 1      & $\pm$1 & $-$1\\
     &   &        &        &        &&        &        &        && $-$1 & $\pm$1 &    3 &$^*$& $-$1   & $\pm$1 &  1 \\ 
   1 & 3 &        &        &        &&        &        &        &&      &        &      && $-$1   & 0      & $-$2 &$^*$ \\
     &   &        &        &        &&        &        &        &&      &        &      &&    1   & 0      &  2   &$^*$ \\
   2 & 1 &     0  & $\pm$2 &  0     && $\pm$1 & $-$1   & $-$2   && 0    & $\pm$2 & 0    && $-$2   & $\pm$1 & $-$1\\
     &   &        &        &        && $\pm$1 & 1      & 2      &&      &        &      && 2      & $\pm$1 & 1  \\
     &   &        &        &        && 0      & 2      & $-$2   &&      &        &      &&        &        & \\
     &   &        &        &        && 0      & $-$2   & 2      &&      &        &      &&       &         & \\  
   2 & 2 & $\pm$1 & 0      & $\pm$1 &&        &        &        && 4    &  0     & $-$2 && $-$2   & 0      & $\pm$ 2\\
     &   &        &        &        &&        &        &        && $-$4 &  0     & 2    &&        &        & \\
   2 & 3 &        &        &        &&        &        &        && $-$2 &  0     & $-$4 &&        &        & \\
     &   &        &        &        &&        &        &        &&    2 &  0     & 4    &&        &        & \\
 \hline \hline
 \end{tabular}
 \end{center}
 \caption{  The assignment of reciprocal lattice vectors to the mode indices $m,n$ for the first two main shells
  and for the orientations used (except [100]). $K_{x[y,z]}$ are given in units of $2\pi/a_{x[y,z]}$ where
  $a_{x[y,z]}$ are the side lengths of the cuboid unit cells of Fig.~\ref{fig:faces} with the desired interface
  orientation in $z$--direction. Within a shell $m=$ const., the RLV are sorted with increasing $|K_z|$,
  for simplicity $n$ and $\bar n$ have been identified. Marked with an asterisk ($^*$) are 
  the monotonously varying leading modes 
  shown in Fig.~\ref{fig:leading_density_modes} below. }
 \label{tab:pmn_all_orient}
\end{table}

\begin{figure}
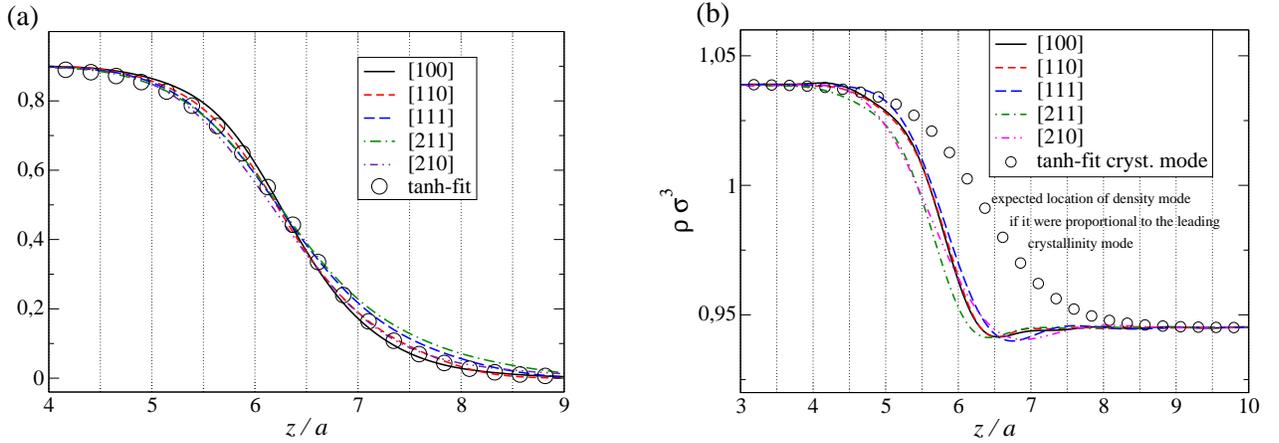

 \vspace*{1cm}
 \epsfig{file=rho_leadingmode_faces.eps, width=7.5cm} \hspace{1.0cm}
 \epsfig{file=rho_avdensitymode_faces.eps , width=8cm}
 \caption{(color online) (a) Leading mode (in units of $\sigma^{-3}$ for all five orientations and a fit of the form 
 $p_{\rm fit}(z)=0.45(1+\tanh[(z-z_0)/(w\,a)])$ with $z_0/a=6.35$ and width $w=1.0$. 
 (b) Average density mode for all 5 orientations,
 for comparison the fitted leading mode, rescaled and shifted to match the asymptotic densities, is given.   }
 \label{fig:leading_density_modes}
\end{figure}

We also found that for the other orientations there is always one leading mode ($m=1$) which shows the monotonous decay
of the leading mode for the [100] orientation, for the associated RLV see Table~\ref{tab:pmn_all_orient}. 
Also the shape is roughly similar for all orientations, and can be roughly
fitted to a tanh--profile with width $a$ (see Fig.~\ref{fig:leading_density_modes} (a)). Also the average
density mode qualitatively has the same shape (especially the depletion zone) for all orientations, see
Fig.~\ref{fig:leading_density_modes} (b). Therefore our previous finding holds regarding the 
clear separation between 
interface position, as determined from the leading crystallinity mode, and the interface position, as determined from the average
density mode. This separation is approximately of size $a \approx 1.6$ $\sigma$ which is not small. So,
coming from the liquid side, the hard sphere fluid first orders and then densifies, in contrast to the
time sequence in homogeneous nucleation where nuclei form by first densifying and then ordering \cite{Schil10}. 

Since the interparticle interaction is short--ranged, one would expect an (oscillatory) exponential decay of the modes 
into the respective bulk phases. An analysis within Landau theory predicts an upper bound of the
decay length of mode $p_{mn}(z)$ of $1/|(\vect K_{\perp,j})_{mn}|$ \cite{Lip89}. A more refined analysis taking 
into account the direct correlation function of the bulk liquid \cite{Loe90,Mik90} also finds that the decay lengths 
decrease with increasing modulus $1/|(\vect K_{\perp,j})_{mn}|$ but the magnitude is enhanced by a factor
of 2\dots 3. We defer a detailed analysis of the decay lengths to future work and merely remark that 
the aforementioned simple rules on the decay lengths do not seem to capture the qualitative importance of modes in the approach to
the bulk phases, see Fig.~\ref{fig:modes_100}. (For example, $p_{00}$ and $p_{22}^+$ should have the same decay
length but the corresponding amplitudes are very different in the approach to the liquid phase. Furthermore, the
different $z$--position of the mode kinks is very important for the relative importance of the modes at a given
location on the $z$--axis.)   

In view of all the features we have seen in Figs.~\ref{fig:modes_100}--\ref{fig:leading_density_modes}
and discussed above, it is clear that restricted density parametrizations can not be very reliable. This has
been shown already for the hard sphere crystal--liquid interface evaluated with a simpler functional
of weighted density type. Full minimization \cite{Ohn94} reduces the surface tension by a factor of  2
compared to a parameterization \cite{Cur87}. Also for FMT functionals quite elaborate parameterizations
have been tested in Ref.~\cite{Son06}. The average surface tension is still larger by about 30\% than the
results given in Table \ref{tab:cr} and the ordering $\gamma_{[100]}>\gamma_{[110]}>\gamma_{[111]}$
is not reproduced.

\subsection{Free energy modes}

\begin{figure}
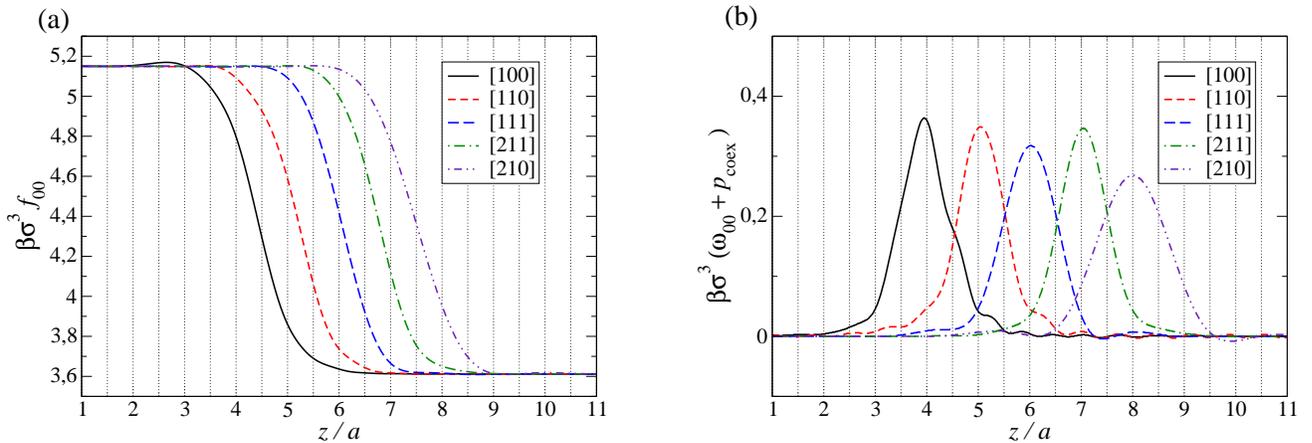

 \vspace*{1cm}
 \epsfig{file=f_faces.eps, width=8cm} \hspace{1.0cm}
 \epsfig{file=om_faces.eps , width=8cm}
 \caption{(color online) Leading free energy modes. (a) Free energy density. (b) Grand free energy density, shifted
 by the coexistence pressure.
 }
 \label{fig:f_om_modes}
\end{figure}

The primary focus of the analysis in the last section was on the density modes which are observable
quantities (see also Sec.~\ref{sec:summary} below). Nevertheless, the mode analysis can also be performed
for the free energy density $f$ or the grand free energy density $\omega$ using the DFT results. The particular
significance of the leading mode $\omega_{00}(z)$ of the grand free energy density is the connection to the
surface tension:
\bea
   \gamma =  \int_{-\infty}^\infty dz (\omega_{00}(z) + p_{\rm coex}) 
\eea
In Fig.~\ref{fig:f_om_modes} we show the leading mode both for the free energy density (a) and
for the grand free energy density (b). The leading free energy modes for the five different orientations
actually look very similar and almost monotonously connect from the higher free energy density of the solid to
the lower free energy density of the liquid (save for a small hump on the crystalline side). It becomes clearer
in the plot for the grand free energy density that there is indeed a broadening of the interface (in free energy terms)
in the sequence [100]-[211]-[110]-[111]-[210]. This is also in rough agreement with the width of the
leading crystallinity modes shown in Fig.~\ref{fig:leading_density_modes} (a).

Modes other than the leading one appear to be without much physical content. However, these may be used to gauge more 
phenomenological approaches such as phase field crystal models \cite{Emm11}.

\section{Summary and Conclusion}

\label{sec:summary}

In this paper, we have presented an analysis of the full 3d density profile of a crystal--liquid interface
in terms of density modes associated with the reciprocal lattice vectors of the bulk crystal.
We have exemplified the extraction of the density modes on finely--meshed density functional data
obtained for the hard sphere system and found a number of interesting results:
\begin{itemize}
 \item a separation of about 1.6 $\sigma$ between  the interface location as determined by the average density
   and the interface location as determined by the leading crystallinity mode
 \item a density depletion zone just in front of the bulk crystal
 \item strongly non-monotonous mode profiles also for next--to--leading modes 
\end{itemize} 
We suggest that simulation data on crystal--fluid interfaces should be analyzed for these modes. 
According to Eq.~(\ref{eq:pmn_extract}), the Fourier transform $\tilde p_{mn}(k_z)$ of the mode $(mn)$
is obtained by "cutting out" the $k_z$--component of the Fourier transform of the 3d density profile  
$\tilde\rho(k_x,k_y,k_z)$ around the reciprocal lattice vector $\vect k = (\vect K_j)_{mn}$
(different $j$ for an index pair $(mn)$ signal mode degeneracy). Practically, this means just averaging 
$\langle \exp(\imag  (\vect K_{\perp,j})_{mn}\cdot \vect r_\perp) \rangle$ in the $x$--$y$--plane
as a function of $z$.
The average would guarantee that the statistics will be good enough for standard runs that one obtains smooth 
Fourier transform and can perform the extraction procedure. Alternatively, also the Fourier transform in
$z$--direction can be sampled directly. This procedure appears to be less prone to statistical uncertainty than 
sampling a full 3d profile.

In principle, the Fourier transforms of the modes $\tilde p_{mn}(k_z)$ are observable quantities in scattering
experiments. Practically, one would need to irradiate a volume containing crystal, liquid and the interface, 
fix the lateral momentum transfer to a particular $(\vect K_{\perp,j})_{mn}$ and finely scan the momentum
transfer in $z$--direction. In view of additional effects like incoherent background and finiteness of the
volume with a well-defined interface orientation, this seems to be a very demanding task. Previous studies
on surface melting using LEED \cite{Pri90} for instance only obtained some integral information on the modes by recording an 
smaller Debye--Waller factor (enhanced decrease in peak intensity) for a certain reciprocal lattice vector due to the decay of
the associated mode. The theoretical analysis done at that time was very rough and could be improved using the 
techniques presented here.

It is perhaps easier to use confocal microscopy for the observation of colloidal crystals \cite{San11,Ege08,Schal11} 
and to do a planar averaging as described for the analysis  
of simulation data. In fact, confocal studies on crystals have previously frequently employed techniques which are borrowed
from simulation such as the bond--order analysis. As remarked in the Introduction, observables connected to bond order
are difficult to handle theoretically, so the mode analysis of simulation and confocal experiments will be beneficial   
for theory development.

{\bf Acknowledgment:} The author thanks the DFG (German Research Foundation) for support through
the Collaborative Research Center SFB-TR6, project N1, and the Priority Program SPP 1296 
(grants SCHI 853/2-2 and OE 285/1-3).


\end{document}